\magnification 1200
\centerline {\bf  $W^{\star}$ Dynamics of Infinite Dissipative Quantum Systems}
\vskip 0.3 cm
\centerline {{\bf by Geoffrey L. Sewell}\footnote*{e-mail address: 
g.l.sewell@qmul.ac.uk}}
\vskip 0.3cm
\centerline {\bf School of Physics and Astronomy, Queen Mary University of 
London,}
\vskip 0.2cm
\centerline {\bf Mile End Road, London E1 4NS, UK}
\vskip 0.5cm
\centerline {\bf Abstract}
\vskip 0.3cm
We formulate the dynamics of an  infinitely extended open dissipative quantum 
system, ${\Sigma}$, in the Schroedinger picture. The generic model on which this is 
based comprisies a $C^{\star}$-algebra, ${\cal A}$, of observables, a folium, ${\cal 
F}$, of  states on this algebra and a one-parameter semigroup,
${\tau}$, of linear transformations of ${\cal F}$ that represents its dynamics and is 
given by a natural infinite volume limit of the corresponding semigroup for a finite 
system.. On this basis, we establish that the dynamics of ${\Sigma}$ is given by a one 
parameter group of completely positive linear transformations of the $W^{\star}$-
algebra dual to ${\cal F}$.This result serves to extend our earlier formulation [1]  of 
infinitely extended conservative systems to open dissipative ones. 
\vskip 0.5cm
\centerline {\bf 1. Introduction.} 
\vskip 0.3cm
The dynamics of a finite dissipative quantum system has been formulated by Lindblad 
[2] and Gorini, Kossakowski and Sudarshan [3] as a one parameter semigroup of 
completely positive (CP) linear transformations of its observables. The aim of this 
article is to provide a corresponding formulation of infinitely extended dissipative 
quantum systems, which may provide a natural basis for a treatment of 
nonequilibrium statistical thermodynamics. We remark that, for the case where these 
reduce to conservative systems, such a treatment has been made [1] for the evolution 
of a folium, ${\cal F}_{\rm con}$ of states, which was shown to be governed by the 
action of a one parameter group of $^{\star}$-automorphisms of the $W^{\star}$-
algebra dual to ${\cal F}_{\rm con}$.. The essential result of the present article is 
that, in the general dissipative case, the quantum evolution of a folium, ${\cal F}$, of 
states is given instead by a one parameter semigroup of completely positive identity 
preserving linear transformations of the $W^{\star}$-algebra dual to ${\cal F}$: in 
general these are quite different, due to their dissipative character, from the 
$^{\star}$-automorphisms of the conservative case.
\vskip 0.2cm
We base our treatment of ${\Sigma}$ on a generic operator  algebraic model of an 
infinitely extended open, dissipative quantum system, as represented by a triple  
$\bigl({\cal A}, \ {\cal F}, \ {\tau}\bigr)$ where ${\cal A}$ is a $C^{\star}$-algebra 
of observables, ${\cal F}$ is a folium\footnote{**}{Recall that [4] a folium is defined 
as a norm closed subset of states that is also closed with respect to convex 
combinations and modifications of the form 
${\omega}{\rightarrow}{\omega}\bigl(B^{\star}(.)B\bigr)/{\omega}(B^{\star}B)$, 
as $B$ runs though
${\cal A}$.} of its states  and ${\tau}$ is a one parameter semigroup of 
transformations of ${\cal F}$ as given by an infinite volume limit of the dynamical 
semigroup of a finite version of ${\Sigma}$.The system is thus an infinite volume 
counterpart of the finite model formulated in [2] and [3]., and is designed to be 
applicable to statistical mechanics and quantum field theory. 
\vskip 0.2cm
We provide further specifications of the above model in subsequent Sections. Thus, in 
Section 2 we pass to a formulation of the algebra ${\cal A}$ in terms of a standard 
quasi-local structure, and in Section 3 we provide corresponding specifications of the 
folium ${\cal F}$ and the dynamical semigroup ${\tau}$. This leads to our main 
result, namely the Proposition of Section 3, which establishes and identifies the 
$W^{\star}$-dynamics of the model.
\vskip 0.5cm 
\centerline {\bf 2. The Algebraic Structure.}. 
\vskip 0.3cm 
We assume that ${\Sigma}$ occupies a space $X$, which may be either ${\bf 
R}^{d}$ or ${\bf Z}^{d}$, with $d$ finite. We denote by $L$  the set of bounded 
open subregions of $X$ and  to each ${\Lambda}{\in}L$ we assign a $W^{\star}$-
algebra, ${\cal A}({\Lambda})$, whose self-adjoint elements represent the bounded 
observables localised in that region. We assume that the algebras 
${\lbrace}{\cal A}({\Lambda}){\vert}{\Lambda}{\in}L{\rbrace}$ are type I factors 
that satisfy the standard requirements of isotony and local commutativity. We define 
${\cal A}_{L}$, the algebra of local observables of ${\Sigma}$, to be 
${\bigcup}_{{\Lambda}{\in}L}{\cal A}({\Lambda})$; and we define ${\cal A}$, the 
norm completion of ${\cal A}_{L}$, to be the $C^{\star}$-algebra of quasi-local 
bounded observables of the system. We assume that each of the local algebras ${\cal 
A}({\Lambda})$ is equipped with a one parameter semigroup 
${\lbrace}{\gamma}_{t}({\Lambda}){\vert}t{\in}{\bf R}_{+}{\rbrace}$ of 
completely positive (CP) [2, 5]] identity preserving transformations, which represent 
the dynamics of the finite version, ${\Sigma}({\Lambda})$, of ${\Sigma}$ confined 
to the region ${\Lambda}$. 
\vskip 0.5cm
\centerline {\bf 3. The Folium ${\cal F}$ and the dynamical semigroup ${\tau}$.}
\vskip 0.3cm
It follows from these specifications [4] that the linear span, $[{\cal F}]$, of ${\cal F}$ 
is the predual of the bicommutant of a certain representation\footnote*{Specifically, 
${\pi}$ is any element of the quasi-equivalence class of the direct sum of the  GNS 
representations of the states comprising ${\cal F}$.}, ${\pi}$, of ${\cal A}$, i.e. that 
${\cal F}$ is the set of normal states on ${\pi}({\cal A})^{{\prime}{\prime}}$. We 
assume that the dynamics of ${\Sigma}$, in the Schroedinger representation, is given 
by a one parameter semigroup, ${\tau}$, of affine transformations of ${\cal F}$. 
Hence, by duality, this semigroup induces a corresponding one, ${\tau}^{\star}$, of 
affine transformations of 
${\pi}({\cal A})^{{\prime}{\prime}}$, as defined by the formula
$${\langle}f;{\tau}_{t}^{\star}B{\rangle}={\langle}{\tau}_{t}f;B{\rangle} \ 
{\forall} \ f{\in}[{\cal F}], \ B{\in}{\pi}({\cal A})^{{\prime}{\prime}}, \
t{\in}{\bf R}_{+},\eqno(3.1)$$
where $[{\cal F}]$ is the linear span of ${\cal F}$. We assume that ${\tau}^{\star}$ 
is just that canonically induced by the local semigroup 
${\lbrace}{\gamma}_{t}({\Lambda}){\vert}t{\in}{\bf R}_{+}{\rbrace}$ in the limit 
${\Lambda}{\uparrow}X$, i.e. 
$${\tau}_{t}^{\star}{\pi}(A)=s:{\rm lim}_{{\Lambda}{\uparrow}X}{\pi}
\bigl({\gamma}_{t}({\Lambda})A\bigr) \ {\forall} \ A{\in}{\cal A}_{L}, \ t{\in}
{\bf R}_{+}.\eqno(3.2)$$
Equivalently, defining 
${\tilde {\gamma}}_{t}({\Lambda})$ to be the transformation of 
${\pi}\bigl({\cal A}({\Lambda})\bigr)$ given by the formula
$${\tilde {\gamma}}_{t}({\Lambda}){\pi}(A):=
{\pi}\bigl({\gamma}_{t}({\Lambda})A\bigr) \ 
{\forall} \ A{\in}{\cal A}_{L}, t{\in}{\bf R}_{+},\eqno(3.3)$$
the condition (3.2) may be expressed in the form 
$${\tau}_{t}^{\star}{\pi}(A)=s:{\rm lim}_{{\Lambda}{\uparrow}X}
{\tilde {\gamma}}_{t}({\Lambda}){\pi}(A) \ {\forall} \ A{\in}{\cal A}_{L}, \ t{\in}
{\bf R}_{+}.\eqno(3.4)$$
\vskip 0.2cm
Suppose now that ${\cal M}$ is a finite dimensional matrix algebra. Then since any 
element, $C$, of ${\pi}({\cal A}_{L}){\otimes}{\cal M}$ may be expressed in the 
form ${\sum}_{r}{\pi}(A_{r}){\otimes}M_{r}$, where the $M_{r}$\rq s form an 
operator valued basis in ${\cal M}$ and the $A_{r}$\rq s are elements of 
${\pi}({\cal A}_{L})$, it follows that the condition (3.4) implies that
$$[{\tau}_{t}^{\star}{\otimes}I]C=s:{\rm lim}_{{\Lambda}{\uparrow}X}
[{\tilde {\gamma}}_{t}({\Lambda}){\otimes}I]C\ {\forall} \ C{\in}
{\pi}({\cal A}_{L}){\otimes}{\cal M}, \ t{\in}{\bf R}_{+}.\eqno(3.5)$$
\vskip 0.3cm
{\bf Proposition.} {\it Under the above specifications of the model,, the  action of the 
dynamical semigroup ${\tau}^{\star}$ on the algebra 
${\pi}\bigl({\cal A}\bigr)^{{\prime}{\prime}}$ is completely positive and idenity 
preserving.} 
\vskip 0.3cm
{\bf Lemma.} {\it Given $(t,{\Lambda}) \ ({\in}{\bf R}_{+}{\times}L)$ the 
transformation ${\tilde {\gamma}}_{t}({\Lambda})$ of 
${\pi}\bigl({\cal A}){\Lambda}\bigr)$ is CP and identity preserving..} 
\vskip 0.3cm
{\bf Proof of Lemma.} Since ${\cal A}({\Lambda})$ is a primary $W^{\star}$-
algebra, it follows from Krauss\rq s formula [6] that the action of 
${\gamma}_{t}({\Lambda}))$ on this algebra may be expressed in the form
$${\gamma}_{t}({\Lambda})C={\sum}_{n{\in}{\bf N}}W_{n}^{\star}CW_{n} \ 
{\forall}C{\in}{\cal A}({\Lambda})\eqno(3.6),$$ 
where ${\lbrace}W_{n}{\rbrace}$ is a sequence of elements of ${\cal 
A}({\Lambda})$ such that
$${\sum}_{n{\in}{\bf N}}W_{n}^{\star}W_{n}=I\eqno(3.7)$$
and ${\sum}_{n{\in}{\bf N}}$ is taken to be the strong limit in the case where the 
number of terms is infinite. Hence, by the normality of ${\pi}$ and Equs. (3.3) and 
(3.6), 
$${\tilde {\gamma}}_{t}({\Lambda}){\pi}(C)=
{\sum}_{n{\in}{\bf N}}{\tilde W}_{n}^{\star}{\pi}(C){\tilde W}_{n}\eqno(3.8)$$
where 
$${\tilde W}_{n}:={\pi}(W_{n})\eqno((3.9)$$ 
and
$${\sum}_{n{\in}{\bf N}}{\tilde W}_{n}^{\star}{\tilde W}_{n}=I.\eqno(3.10)$$
Now let ${\cal M}$ be a finite dimensional matrix algebra. Then any element 
${\tilde C}$ of ${\pi}\bigl({\cal A}({\Lambda})\bigr){\otimes}{\cal M}$ may be 
expressed as a finite sum
$${\tilde C}={\sum}_{r{\in}J}{\pi}(C_{r}){\otimes}M_{r}\eqno(3.11),$$
where $J$ is a finite index set, the $C_{r}$\rq s are elements of 
${\cal A}({\Lambda})$ and the $M_{r}$\rq s form an operator basis for ${\cal M}$. 
Hence, by Equs. (3.8)-(3.11),
$$[{\tilde {\gamma}}_{t}({\Lambda}){\otimes}I]({\tilde C}^{\star}{\tilde C})=
[{\tilde {\gamma}}_{t}({\Lambda}{\otimes}I]
{\sum}_{r,s{\in}J}{\pi}(C_{r}^{\star}C_{s}){\otimes}
M_{r}^{\star}M_{s}=$$
$${\sum}_{r,s{\in}J;,n{\in}{\bf N}}
{\tilde W}_{n}^{\star}{\pi}(C_{r}^{\star}C_{s}){\tilde W}_{n}
{\otimes}M_{r}^{\star}M_{s}=
{\sum}_{n{\in}{\bf N}}D_{n}^{\star}D_{n}\eqno(3.12),$$
where
$$D_{n}={\sum}_{r{\in}J}{\pi}(C_{r})
{\tilde W}_{n}{\otimes}M_{r}.\eqno(3.13)$$
Hence ${\tilde {\gamma}}_{t}({\Lambda}){\otimes}I$ is positive and therefore 
${\tilde {\gamma}}_{t}({\Lambda})$ is CP. Further, by Eqs. (3.8) and (3.10), it is 
identity preserving.
\vskip 0.3cm
{\bf Proof of Proposition.} It follows immediately from the Lemma, together with  the 
definition (3.3) of  ${\tilde {\gamma}}_{t}({\Lambda})$ and the complete positivity 
of ${\gamma}_{t}({\Lambda})$, that the transformation ${\tilde 
{\gamma}}_{t}({\Lambda})$ is both CP and identity preserving. 
\vskip 0.2cm
To prove that the same is true for ${\tau}_{t}^{\star}$, we first infer from Eqs. (3.3) 
and (3.4) that its identity preserving property follows from that of 
${\gamma}_{t}({\Lambda})$. 
\vskip 0.2cm
Next we note that, by the finite dimensionality of ${\cal M}$, elements ${\tilde B}$ 
of ${\pi}({\cal A}_{L}){\otimes}{\cal M}$ take the form
$${\tilde B}={\sum}_{J}{\pi}(B_{J}){\otimes}M_{J}$$
where $J$ is a finite index set. Hence
$$[{\tau}_{t}^{\star}{\otimes}I]{\tilde B}=
{\sum}_{J}{\tau}_{t}^{\star}{\pi}(B_{J}){\otimes}M_{J}$$
a.nd hence, by Equ. (3.5),
$$[{\tau}_{t}^{\star}{\otimes}I]{\tilde B}=
s-{\lim}_{{\Lambda}{\uparrow}X}
\bigl({\tilde{\gamma}}_{t}({\Lambda}){\otimes}I\bigr){\tilde B}$$
Therefore
$$[{\tau}_{t}^{\star}{\otimes}I]({\tilde B}^{\star}{\tilde B})=
s-{\lim}_{{\Lambda}{\uparrow}}{\tilde {\gamma}}_{t}({\Lambda})
({\tilde B}^{\star}{\tilde B})$$
Since it follows from the Lemma that ${\tilde {\gamma}}_{t}({\Lambda})$ is CP,  
i.e. that ${\tilde {\gamma}}_{t}({\Lambda}){\otimes}I$ is positive, it follows from 
the last equation that the same is true for ${\tau}_{t}^{\star}{\otimes}I$, for all 
values of the dimensionality of ${\cal M}$. In other words, ${\tau}^{\star}$ is CP.
\vskip 0.5cm
\centerline {\bf References}
\vskip 0.3cm\noindent
[1] G. L. Sewell: Lett. Math. Phys. {\bf 6}, 209-213 (1982)
\vskip 0.2cm\noindent
[2] G. Lindblad: Commun. Math. Phys. {\bf 48}, 119-130 (1976)
\vskip 0.2cm\noindent
[3] V. Gorini, A. Kossakowski and E. C..G. Sudarshan: J. Math. Phys. {\bf 17}, 821-5 
(1976)
\vskip 0.2cm\noindent
[4] R. Haag, R V. Kadison and D. Kastler: Commun. Math. Phys.{\bf 33}, 1-22 
(1973)
\vskip 0.2cm\noindent
[5] W. F. Stinespring: Proc. Amer. Math. Soc. {\bf 6}, 211-216 (1955)
\vskip 0.2cm\noindent
[6] K. Kraus: Ann. Phys. {\bf 64}, 311-335 (1971) 

\end